\documentclass[pdflatex,iicol,sn-mathphys-num]{sn-jnl}
\usepackage{amsmath,amssymb}
\usepackage{graphicx}
\renewcommand{\orcidlogo}{%
  \raisebox{-0.15ex}{\includegraphics[height=1.35ex]{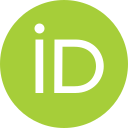}}}

\begin{document}

\title[Discovery sensitivity with background uncertainty]
{Discovery Sensitivity for a Counting Experiment with Background Uncertainty}

\author*[1]{\fnm{Enzo} \sur{Canonero}\,\orcid{https://orcid.org/0000-0002-7180-4562}}\email{Enzo.Canonero@rhul.ac.uk}
\author*[1]{\fnm{Glen} \sur{Cowan}\,\orcid{https://orcid.org/0000-0001-8363-9827}}\email{G.Cowan@rhul.ac.uk}

\affil[1]{\orgdiv{Physics Department},
  \orgname{Royal Holloway, University of London},
  \orgaddress{\city{Egham}, \postcode{TW20 0EX}, \country{UK}}}

\abstract{
%
In Particle Physics, a search for a new signal process is often
based on observing a Poisson-distributed number of events,
whose mean contains contributions from background and, if it 
exists, the hypothesised signal.   The discovery significance 
can be expressed as an equivalent
number of standard deviations derived from the $p$-value of the
background-only hypothesis.   To characterise the experimental 
sensitivity, one may report the median, assuming a nominal signal
strength, of the discovery significance.
In this paper, approximate expressions for the median significance 
are derived both when the expected number of background events is
known and when the background rate is uncertain but constrained by 
a Poisson control measurement.
%
%
The formulae are based on a test statistic using the profile
likelihood ratio, and the median significance is approximated using
the Asimov data set. Higher-order asymptotic corrections, based on the
Barndorff-Nielsen $r^\ast$ statistic, are incorporated for both the 
observed and expected discovery significance.
The validity of the resulting expressions is compared with Monte Carlo
results and with other formulae for expected significance often used in
particle physics. The higher-order corrections are found to provide 
meaningful improvements at small event yields.  The results are important 
for obtaining an accurate assessment of the sensitivity of a planned 
experiment and for the optimal choice of cuts that determine the expected numbers of signal and background events.
}


\maketitle

\section{Introduction}
\label{sec:intro}

In a search for a new particle-physics process, one may observe a
certain number of events, $n$, modelled as following a Poisson
distribution with a mean of $s + b$, where $s$ and $b$ represent the
expected numbers of events from the signal process and
background processes, respectively.  Here this will be referred to as
a Poisson counting experiment. To establish a discovery of the
signal, one reports the $p$-value of the background-only ($s =
0$) hypothesis or, equivalently, the Gaussian significance $Z$.
This is related to the $p$-value by

\begin{equation}
\label{eq:Zval}
Z = \Phi^{-1}(1 - p) \;,
\end{equation}

\noindent where $\Phi^{-1}$ is the quantile function (inverse cumulative
distribution function) of the standard Gaussian distribution. When the
$p$-value refers to the background-only hypothesis, we
refer to the corresponding $Z$ value as the discovery significance.
In particle physics, a widely used threshold for a discovery has been a
$p$-value of $2.9 \times 10^{-7}$ or less, corresponding to a
significance of $Z = 5$ or more.

When designing a new experiment, it is important to know what discovery
significance to expect if a certain signal model is in fact true.  For
this one can report the mean or median value of $Z$ under the assumption
of some nominal value of $s$, i.e., assuming that $n$ will have a mean
of $s+b$.  As noted in Ref.~\cite{asimov}, because the $p$-value and
significance $Z$ have a nonlinear, monotonic relation, it is
convenient to use the median rather than the expectation value of the significance,
so that the median $Z$ is given by $Z$ evaluated with the median $p$.

Often the expected number of background events $b$ is uncertain and must
be treated as an adjustable parameter.  To constrain its value, one 
may carry out a control measurement that yields a Poisson-distributed 
value $m$ with mean $\tau b$, where $\tau$ is a known scale factor.  This 
problem is discussed in Sec.~\ref{sec:bunknown}. The first major result 
of this paper is the extension of the formula for the median significance
to the case where $b$ is uncertain, as given in Eq.~(\ref{eq:onoffmedZ2})
of Sec.~\ref{sec:bunknown}. This formula was
included in a conference presentation~\cite{bib:Cowan2012} in 2012 but not
published at that time,  although it has been used in a number of studies,
e.g.,~\cite{bib:Kumar2015,bib:Bhattiprolu2021}.  
An important purpose of the present paper  is to provide a stable 
reference for this result.

The results mentioned above use asymptotic distributions that stem from 
theorems of Wilks~\cite{bib:Wilks1938} and Wald~\cite{bib:Wald1943}, and
they are accurate when the expected event yields are sufficiently large.
At small event yields, however, these approximations can incur 
non-negligible errors.  The second important focus of the paper is the 
application of higher-order asymptotic corrections based on the 
Barndorff-Nielsen
$r^\ast$ statistic~\cite{bib:Barndorff-Nielsen1980,
bib:Barndorff-Nielsen1983,
bib:Barndorff-Nielsen1986,
bib:Barndorff-Nielsen1990,
bib:Barndorff-Nielsen1991}
to both the observed and median discovery significance.
In this way, we extend the validity of the results to the case of even lower
event counts, as shown in Sec.~\ref{sec:hoa}.
Conclusions are given in Sec.~\ref{sec:conclusions}.

\section{Discovery as a statistical test}
\label{sec:disctest}

In particle physics, one frequently quantifies the significance of an
observed signal by quoting the $p$-value of the background-only
hypothesis, i.e., $s=0$. One method for defining the $p$-value for a
hypothesised value of $s$ is to construct a test statistic based on the
profile likelihood ratio,

\begin{equation}
\label{eq:PLR} 
\lambda(s) = \frac{ L(s, \hat{\hat{\boldsymbol{\theta}}}(s)) } 
{L(\hat{s}, \hat{\boldsymbol{\theta}}) } \;.
\end{equation}

\noindent Here $L(s, \boldsymbol{\theta})$ is the likelihood function for
the measurement (i.e., the number of events $n$ plus any subsidiary
measurements), conditional on the signal parameter $s$ and any additional
(nuisance) parameters $\boldsymbol{\theta}$. A single hat denotes an
unconstrained maximum-likelihood estimator (MLE), whereas the double-hat
notation in the numerator of Eq.~(\ref{eq:PLR}) denotes the value of
$\boldsymbol{\theta}$ that maximises the likelihood for the specified
value of $s$, i.e., the constrained (or profiled) MLE. The numerator is
thus the profile likelihood; the denominator is the maximum likelihood.

In Ref.~\cite{asimov} the profile likelihood ratio was used as the
basis of a test statistic defined as

\begin{equation}
\label{eq:q0} 
q_{0} =  
\left\{ \! \! \begin{array}{ll}
               - 2 \ln \lambda(0)  
               & \quad \hat{s} \ge 0 \;, \\*[0.3 cm]
               0 & \quad \hat{s} < 0  \;.
              \end{array}
       \right.
\end{equation}

\noindent This is defined so that large values of $q_0$ correspond to
increasing disagreement between the data and the hypothesis $s=0$.
Although we consider here only the case where a physical
signal model has $s > 0$, the estimator $\hat{s}$ is defined as the
value that maximises the likelihood even if this is negative.  This
occurs if the number of observed events is smaller than the expected
background.  The statistic $q_0$ is defined as zero for $\hat{s} < 0$
so that it reflects a discrepancy between the data and the hypothesis only
in the case where the observed signal rate is positive.

\section{\texorpdfstring{$s/\sqrt{b}$}{s/sqrt(b)} and related measures of discovery sensitivity}
\label{sec:soverrootb}

In particle physics the quantity $s/\sqrt{b}$ has been widely used as
a measure of expected discovery significance.  The rationale behind
this formula is that a Poisson-distributed quantity $n$ with a large
mean value $s+b$ can be approximated by a Gaussian-distributed
variable $x$ with mean $s+b$ and standard deviation $\sqrt{s+b}$.  The
$p$-value of the background-only hypothesis given an observation $x$
is therefore

\begin{equation}
\label{eq:pvalx}
p = 1 - \Phi \left( \frac{x - \mu}{\sigma} \right) = 
1 - \Phi \left( \frac{ x - b }{\sqrt{b}} \right) \;,
\end{equation}

\noindent where $\mu = b$ and $\sigma = \sqrt{b}$ refer to the mean
and standard deviation of $x$ under the assumption of $s=0$.  Using this
with Eq.~(\ref{eq:Zval}) gives the discovery significance

\begin{equation}
\label{eq:Zx}
Z = \frac{x -b}{\sqrt{b}} \;.
\end{equation}

\noindent The median (here equal to the mean) $Z$ under the assumption of
a given value of $s$ is therefore

\begin{equation}
\label{eq:soverrootb}
\mbox{med}[Z | s] = \frac{s}{\sqrt{b}} \;.
\end{equation}

\noindent The intuitive explanation of this formula is that the
standard deviation of $n$ assuming background only is $\sqrt{b}$, and
therefore the ratio $s/\sqrt{b}$ represents the size of the signal
divided by the statistical error on $n$ expected under the
background-only hypothesis.

Often the expected number of background events is not known exactly
but has some systematic uncertainty characterised by a standard
deviation $\sigma_b$.  In this case, one may generalise
Eq.~(\ref{eq:soverrootb}) to account for both the statistical and
systematic errors in $b$ by replacing $\sqrt{b}$ by the quadratic
sum of $\sqrt{b}$ and $\sigma_b$, so that the median significance
becomes

\begin{equation}
\label{eq:soverrootbsys}
\mbox{med}[Z | s] = \frac{s}{\sqrt{b + \sigma_b^2}} \;.
\end{equation}

\noindent 
In Sec.~\ref{sec:bunknown} a formal justification for
Eq.~(\ref{eq:soverrootbsys}) will be given and the limits of its
validity investigated.

\section{Poisson case with known background}
\label{sec:bknown}

If the expected number of background events, $b$, is known with
negligible uncertainty, then the likelihood function for
the Poisson counting experiment is

\begin{equation}
\label{eq:poissonL}
L(s) = \frac{(s+b)^n}{n!} e^{-(s+b)} \;.
\end{equation}

\noindent In Ref.~\cite{asimov} this problem was investigated using
the test statistic $q_0$ as defined in Eq.~(\ref{eq:q0}).  For a
sufficiently large expected number of events, one can show, using
Wilks' theorem (see, e.g.,~\cite{asimov} and references therein), that
the discovery significance can be approximated by $Z = \sqrt{ q_0}$.
For the present problem, this gives

\begin{equation}
\label{eq:Z0count}
Z = \sqrt{ 2 \left( n \ln \frac{n}{b} + b - n \right) } \;,
\end{equation}

\noindent for $n > b$ and $Z = 0$ otherwise.  It was also shown in
Ref.~\cite{asimov} that one can approximate the median significance by
replacing the data by the corresponding expectation values (the
so-called Asimov data set).  Substituting $s+b$ for $n$ in
Eq.~(\ref{eq:Z0count}) thus gives the Asimov significance $Z_{\rm A}$,
used as an approximation to $\mbox{med}[Z | s]$:

\begin{equation}
\label{eq:Zmedcount}
Z_{\rm A} = 
\sqrt{ 2 \left( (s+b) \ln \left( 1 + \frac{s}{b} \right) - s \right) } \;.
\end{equation}

\noindent Expanding the logarithm in $s/b$ to second order, one finds

\begin{equation}
\label{eq:Zmedcount2}
Z_{\rm A} = \frac{s}{\sqrt{b}} 
\left( 1 + {\cal O}(s/b) \right)\;. 
\end{equation}

\noindent Thus the full expression for $Z_{\rm A}$ in Eq.~(\ref{eq:Zmedcount})
reduces to the widely used formula $s/\sqrt{b}$ for $s \ll b$.

\begin{figure}[t]
  \centering
  \includegraphics[width=\columnwidth]{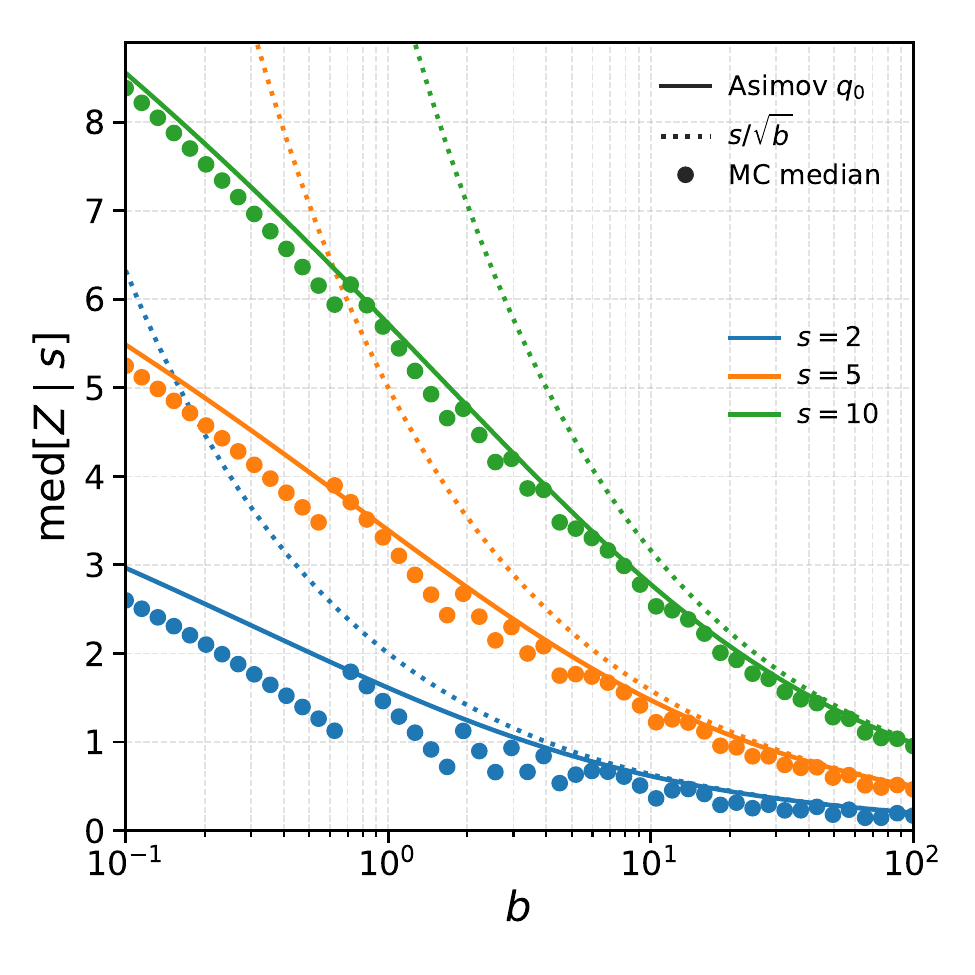}
  \caption{Median discovery significance $\mbox{med}[Z | s]$ as a function
    of the expected background $b$ for the known-background counting
    experiment for $s=2,5$ and $10$. The first-order Asimov significance
    $Z_{\rm A}$ obtained from $q_0$ is compared with the limiting expression
    $s/\sqrt{b}$ and the MC estimate of $\mbox{med}[Z | s]$. This plot
    reproduces the results shown in Fig.~7 of Ref.~\cite{asimov}.}
  \label{fig:medsig}
\end{figure}

Median expected discovery significances $\mbox{med}[Z | s]$ for
$s = 2, 5$ and $10$ are shown in Fig.~\ref{fig:medsig}.  The solid
curves show the Asimov significance $Z_{\rm A}$ of
Eq.~(\ref{eq:Zmedcount}), the dotted curves give the approximation
$s/\sqrt{b}$, and the points are Monte Carlo estimates of
$\mbox{med}[Z | s]$.
The structure seen in the points is due to the discrete
nature of the data.  One sees that Eq.~(\ref{eq:Zmedcount}) provides a
much better approximation to the true median than does $s/\sqrt{b}$ in
regions where $s \ll b$ does not hold.

\section{Poisson case with uncertain background}
\label{sec:bunknown}

If the expected number of background events, $b$, is not known, one
must treat it as a nuisance parameter in the likelihood function.
Because $b$ could be adjusted to accommodate any observed number of
events, it would be impossible to reject the hypothesis of $s=0$
unless some additional information is introduced that constrains $b$.
Often this is done using a control measurement that counts the
number of events $m$ in a data sample in which signal events are believed
to be absent. The mean control count can be related to $b$, the expected
number of background events in the primary measurement of $n$.
For example, $m$ may be modelled as a Poisson-distributed variable with
mean $\tau b$, where $\tau$ is a scale factor taken here to be known with
negligible uncertainty. 

The full likelihood function that describes both the primary measurement
$n$ and the control measurement $m$ is therefore the product of
the two corresponding Poisson distributions:

\begin{equation}
\label{eq:onoffL}
L(s,b) = \frac{(s+b)^n}{n!} e^{-(s+b)} \, 
\frac{(\tau b)^m}{m!} e^{-\tau b} \;.
\end{equation}

\noindent This problem has been studied in both particle physics and astrophysics (often called the ``on/off'' problem; see, e.g.,~\cite{asimov,LiMa83,Cranmer05,Cousins08}).
To find the profile likelihood ratio, one needs the estimators for $b$
and $s$ as well as the conditional estimator for $b$ given a value of
$s$:

\vspace{-0.5cm}

\begin{align}
\hat{s}
  &= n - m/\tau \;,
  \label{eq:shat}\\[0.2cm]
\hat{b}
  &= m/\tau \;,
  \label{eq:bhat}\\[0.2cm]
\hat{\hat{b}}(s)
  &= \frac{1}{2(1+\tau)}
     \Bigl\{n+m-(1+\tau)s
     \notag\\
  &\quad
     +\sqrt{\bigl(n+m-(1+\tau)s\bigr)^2
     +4(1+\tau)sm}\Bigr\}\;.
  \label{eq:bhathat}
\end{align}

\noindent For the statistic $q_0$ one needs in particular $\hat{\hat{b}}(0)$,
which, from Eq.~(\ref{eq:bhathat}), is

\begin{equation}
\label{eq:bhathat0}
\hat{\hat{b}}(0) = \frac{n + m}{1 + \tau} \;.
\end{equation}

As in Sec.~\ref{sec:bknown}, we use the approximation $Z = \sqrt{q_0}$,
valid in the large-sample limit, which gives

\begin{equation}
\label{eq:onoffZ}
\begin{split}
Z ={}&
\Biggl[-2\Biggl(
n\ln\left[\frac{n+m}{(1+\tau)n}\right]
\\
&\qquad{}
+m\ln\left[\frac{\tau(n+m)}{(1+\tau)m}\right]
\Biggr)\Biggr]^{1/2}\;,
\end{split}
\end{equation}

\noindent for $n > \hat{b}$ and $Z = 0$ otherwise.  This is the same
as Eq.~(17) of Ref.~\cite{LiMa83} and Eq.~(25) of
Ref.~\cite{Cousins08}.

As in Sec.~\ref{sec:bknown}, we replace the data values $n$ and $m$
by their expectation values $s+b$ and $\tau b$ to give the
``Asimov'' approximation for the median significance.  Making
this substitution in Eq.~(\ref{eq:onoffZ}) gives

\begin{equation}
\label{eq:onoffmedZ}
\begin{split}
Z_{\rm A} ={}&
\Biggl[-2\Biggl(
(s+b)\ln\left[
\frac{s+(1+\tau)b}{(1+\tau)(s+b)}
\right]
\\
&\qquad{}
+\tau b\ln\left[
1+\frac{s}{(1+\tau)b}
\right]
\Biggr)\Biggr]^{1/2}\;.
\end{split}
\end{equation}

\noindent The case where the control measurement $m$ has a small
relative statistical uncertainty corresponds to very large $\tau$, and
in this limit Eq.~(\ref{eq:onoffmedZ}) reverts to the expression for known
$b$ given by Eq.~(\ref{eq:Zmedcount}).

It is useful to re-express Eq.~(\ref{eq:onoffmedZ}) in terms of the
background uncertainty inferred from the control measurement $m$.
The estimator for $b$ is given by
Eq.~(\ref{eq:bhat}), and because the variance of $m$ is equal to its
mean, $\tau b$, the variance of $\hat{b}$ is

\begin{equation}
\label{eq:sigma2b}
V[\hat{b}] \equiv \sigma^2_b = \frac{b}{\tau} \;.
\end{equation}

\noindent Using this equation to eliminate $\tau$ from
Eq.~(\ref{eq:onoffmedZ}) gives the result

\begin{equation}
\label{eq:onoffmedZ2}
\begin{split}
Z_{\rm A} ={}&
\Biggl[2\Biggl(
(s+b)\ln\left[
\frac{(s+b)(b+\sigma_b^2)}
     {b^2+(s+b)\sigma_b^2}
\right]
\\
&\qquad{}
-\frac{b^2}{\sigma_b^2}
\ln\left[
1+\frac{\sigma_b^2s}
        {b(b+\sigma_b^2)}
\right]
\Biggr)\Biggr]^{1/2}\;.
\end{split}
\end{equation}

\noindent By expanding this expression in powers of $s/b$ and
$\sigma_b^2/b$, one finds

\begin{equation}
\label{eq:onoffmedZ3}
Z_{\rm A} = \frac{s}{\sqrt{b + \sigma_b^2}} \left( 1 + {\cal O}(s/b)
+ {\cal O}(\sigma_b^2/b) \right) \;.
\end{equation}

One sees that the expression in Eq.~(\ref{eq:soverrootbsys}), originally
justified on intuitive grounds, also follows from calculating the
significance using Wilks' theorem and the Asimov data set. The limiting formula given by
Eq.~(\ref{eq:onoffmedZ3}) is expected to be valid in cases where one
has $s \ll b$ and $\sigma_b^2 \ll b$.  From Eq.~(\ref{eq:sigma2b}) we
have $\sigma_b^2/b = 1/\tau$, and because the expectation value of $m$
is $E[m] = \tau b$, the requirement $\sigma_b^2 \ll b$ is equivalent
to $E[m] \gg b$ or $\tau \gg 1$.  That is, the expected number of
events in the control region should be large compared with the expected
number of background events contributing to the primary measurement of
$n$ (and in addition $s \ll b$ must hold).

\begin{figure*}[p]
  \centering
  \includegraphics[width=0.84\textwidth]{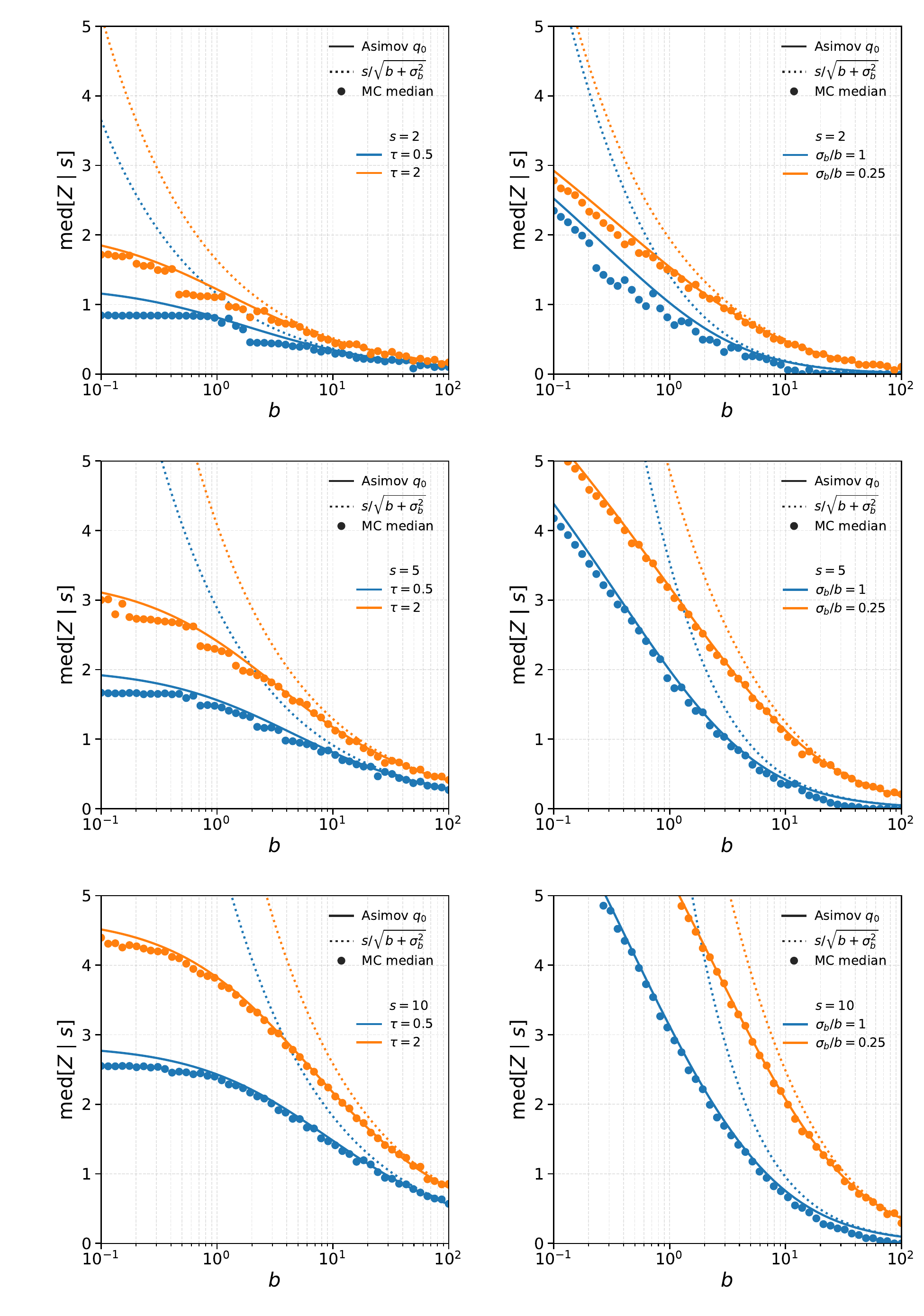}
  \caption{Median discovery significance $\mbox{med}[Z | s]$ as a function
    of $b$ for the uncertain-background model with $s=2$ (top), $s=5$ (middle) and
    $s=10$ (bottom).  The left column fixes $\tau=0.5$ or $2$, and
    the right column fixes $\sigma_b/b=0.25$ or $1$. The first-order
    Asimov significance $Z_{\rm A}$ obtained from $q_0$ is compared with
    $s/\sqrt{b+\sigma_b^2}$ and the MC estimate of $\mbox{med}[Z | s]$.
    MC points that reach
    the $5\sigma$ toy-resolution ceiling are omitted.}
  \label{fig:medsigsys}
\end{figure*}

Figure~\ref{fig:medsigsys} shows the median discovery significance for
$s = 2, 5$ and $10$ as a function of $b$, using two ways of quantifying
the background uncertainty.  The curves in the left-hand panels 
correspond to $\tau = 0.5$ or $2$, i.e., the scale factor between the 
sizes of the primary and control regions is independent of $b$. 
The right-hand panels use fixed relative uncertainty with 
$\sigma_b/b = 0.25$ or $1$.  The MC points are obtained by generating
pseudo-data $(n,m)$ at $(s,b)$ and,
for each pair, estimating the background-only $p$-value by resampling at
$s=0$ with $b$ fixed to its profiled value
$\hat{\hat b}_0=\hat{\hat b}(s=0)$ from Eq.~(\ref{eq:bhathat0}). 
This procedure is known as 
profile construction~\cite{Cranmer05} or hybrid
resampling~\cite{ChuangLai2000,BodhisattvaSen2009}. The structure in
the points is due to the discreteness of the Poisson-distributed data.
The dotted and solid curves show the predictions of
Eqs.~(\ref{eq:soverrootbsys}) and~(\ref{eq:onoffmedZ2}), respectively.
Although both formulae agree with the Monte Carlo values for
sufficiently large $b$, the full formula from
Eq.~(\ref{eq:onoffmedZ2}) is clearly in far better agreement for low
$b$, and the improvement is more pronounced at large $s$.  This is 
understandable because for low $b$ and large $s$
the ratios $s/b$ and $\sigma^2_b/b$ become large and thus the 
approximation of Eq.~(\ref{eq:soverrootbsys}) is expected to break down.

\section{Higher-order asymptotic corrections}
\label{sec:hoa}

The formulae presented above in Eqs.~(\ref{eq:Zmedcount}) and 
(\ref{eq:onoffmedZ2}) for
the median discovery significance rely on the large-sample assumptions
of Wilks' theorem, and one therefore expects them
to break down for small event yields.  As seen in Figs.~\ref{fig:medsig} 
and~\ref{fig:medsigsys}, the approximations remain reasonably
good provided the expected numbers of signal and background
events are at least a few, and this already makes them valuable
tools in many cases of practical interest.  The usable range can 
be extended to even lower event yields, however, by using 
higher-order asymptotic corrections 
(see, e.g.,  Ref.~\cite{Brazzale07}).  Specifically, we use the
Barndorff-Nielsen $r^\ast$ statistic~\cite{
bib:Barndorff-Nielsen1980,bib:Barndorff-Nielsen1983,
bib:Barndorff-Nielsen1986,bib:Barndorff-Nielsen1990, 
bib:Barndorff-Nielsen1991}: we describe its construction and show how 
it improves the accuracy of the observed discovery
significance at low expected counts for both the known- and 
uncertain-background cases.

At low expected event yields, the discrete nature of the 
Poisson-distributed measurements
becomes increasingly relevant. We therefore discuss how to account for
discreteness when approximating Poisson probabilities using a continuous
distribution. Such a correction is naturally defined when the background is
known, but has no unique extension when it is uncertain.

Finally, we study how the higher-order and continuity corrections can be
applied to the Asimov estimate of the median significance.

\subsection{The
Barndorff-Nielsen \texorpdfstring{$r^\ast$}{r*} statistic}
\label{sec:rstar}

To test a hypothesised signal value $s$, we define the signed 
likelihood-ratio
root, i.e., the signed square root of the likelihood-ratio test statistic,

\begin{equation}
\label{eq:r}
r(s) = \mathrm{sign}(\hat{s} - s)\sqrt{-2\ln\lambda(s)} \;.
\end{equation}
This provides the starting point for the higher-order asymptotic
corrections discussed below. Under the regularity conditions of Wilks' 
theorem, $r$ is asymptotically distributed as a
standard Gaussian. With this definition, the discovery test statistic in Eq.~(\ref{eq:q0}) can
equivalently be expressed in terms of the signed likelihood-ratio root as
\begin{equation}
q_0 =
\begin{cases}
r(0)^2 & \quad \hat{s} \geq 0 \;, \\[0.2cm]
0      & \quad \hat{s} < 0 \;.
\end{cases}
\end{equation}
Since $r(0)$ and $\hat{s}$ have the same sign, this can be written more
compactly as
\begin{equation}
q_0 = \left[ \max\{ 0, r(0) \} \right]^2 \;.
\end{equation}
In the approximation that $r(0)$ follows a standard Gaussian, 
the $p$-value of $s=0$ is $p = 1-\Phi(r(0))$, and thus $Z = r(0)$ 
is equivalent  to the result
based on Wilks' theorem used throughout the preceding sections. 
To be consistent with Ref.~\cite{asimov}, we take 
$Z = \sqrt{q_0} = \max\{0,r(0)\}$. 
In this way, the significance is simply set to zero if
the data value $n$ is less than the expected background. This convention
is widely used in particle physics and still gives the correct median 
significance under the assumption of positive $s$.

The Barndorff-Nielsen $r^\ast$ statistic refines the signed 
likelihood-ratio root $r$ so that its sampling distribution is 
more accurately approximated by a
standard Gaussian. It is obtained (see, e.g.,~\cite{Brazzale07})
by replacing $r(s)$ with
\begin{equation}
\label{eq:rstar}
r^\ast(s) = r(s) + \frac{1}{r(s)}
\ln\!\left|\frac{u(s)}{r(s)}\right| \;,
\end{equation}
\noindent where $u(s)$\footnote{The auxiliary statistic is 
denoted by $q$ in many references on the $r^\ast$ correction.
Here we use $u$ to avoid confusion with the discovery test statistic
$q_0$.} is a model-dependent auxiliary statistic found from derivatives
of the log-likelihood with respect to the maximum-likelihood estimator of
the parameter of interest, i.e., $\partial \ln L/ \partial \hat{s}$.  
Specifically,
$u(s)$ is related to the change in this derivative when
evaluated at the constrained and unconstrained estimates of the nuisance
parameter, $\hat{\hat{b}}(s)$ and $\hat{b}$.
For an exact definition of $u$ and details of how it is computed, see,
e.g., Refs.~\cite{Brazzale07,Reid2000,Canonero23}. 

The statistic $u(s)$ is defined such that it approaches $r(s)$ 
and thus $r^{\ast}$ reverts to $r$ in the large-sample limit.
If the large-sample condition does not hold, that is, 
with small Poisson counts, the corrected $r^{\ast}$ leads to
a more accurate approximation of the $p$-value than
does the original $r$.   This is particularly relevant in 
the tails of the distribution, i.e., where
the $p$-value is small, which is the regime of interest for discovery.

The $r^\ast$ replacement gives the corrected discovery
test statistic, which we define as

\begin{equation}
\label{eq:q0star}
q_0^\ast = \left[\max\{0,r^\ast(0)\}\right]^2 \;.
\end{equation}
At first order, a non-zero discovery significance is obtained only when
$\hat{s}>0$, or equivalently when $r(0)>0$. 
With the higher-order correction, the corresponding requirement
is instead applied to $r^{\ast}(0)$ using Eq.~(\ref{eq:q0star}),
and in this way 
only $r^\ast(0)>0$ is interpreted as discovery-like evidence
after accounting for finite-sample and nuisance-parameter effects. In the
large-sample limit one has $r^\ast(0)\rightarrow r(0)$, so the 
usual condition on
$\hat{s}$ is recovered. The corrected significance is therefore
$Z=\sqrt{q_0^\ast}=\max\{0,r^\ast(0)\}$.

In the remainder of this paper we express all results in terms of the
discovery test statistics $q_0$ and $q_0^\ast$, in keeping with the
notation commonly used in particle physics for discovery tests.

\subsection{Continuity correction for discrete data}
\label{sec:continuity}

The $p$-values associated with $q_0$ and $q_0^\ast$ are obtained from
the normal upper tail as $p = 1-\Phi(\sqrt{q_0}\,)$ and
$p = 1-\Phi(\sqrt{q_0^\ast}\,)$, respectively. These are continuous tail
approximations, whereas the counting experiments considered here are based
on discrete Poisson observations. Some care is therefore required when
comparing these approximations with exact Poisson $p$-values. In particular, for a single
Poisson count $N$, it can be
shown that the normal-tail approximation based on either $q_0$ or
$q_0^\ast$ approximates the \textit{mid $p$-value}~\cite{DavisonEtAl2006,Lancaster1961},

\begin{equation}
\label{eq:mid-p}
p_{\rm mid}(n)
=
P(N>n)+\frac12 P(N=n) \;.
\end{equation}
This follows because, at the observed count, the continuous approximation
passes through the midpoint of the discrete cumulative distribution's jump
of size $P(N=n)$.


By convention, however, the $p$-value is taken to include
the probability of the observed number of counts, i.e.,

\begin{equation}
p(n)=P(N\ge n) \;.
\end{equation}
To approximate this inclusive Poisson tail with a continuous statistic,
the observed count is associated with its unit bin and the statistic is
evaluated at the lower bin edge $n\rightarrow n-\tfrac{1}{2}$. 
Accordingly, for the known-background
counting experiment the continuity correction is applied to $q_0^\ast$ as

\begin{equation}
\label{eq:rstarcc}
q_0^\ast(n)
\;\rightarrow\;
q_0^\ast\!\left(n-\tfrac{1}{2}\right) \;.
\end{equation}
In principle, the same prescription could also be applied to the
first-order statistic $q_0$. In this paper, however, we retain
the conventional definition of $q_0$ evaluated at the observed integer
count and apply the continuity correction only to $q_0^\ast$. This keeps
the first-order approximation identical to the standard result widely
used in particle physics.

For multidimensional discrete problems, such as the uncertain-background
model of Sec.~\ref{sec:bunknown}, the observation is a discrete 
pair $(n,m)$, and so the
discovery region is no longer defined by a threshold on a single variable.
There is therefore no unique analogue of the half-bin shift.
Accordingly, throughout this paper we apply the continuity correction 
only in the known-background counting problem, and then only 
together with the higher-order correction.

\subsection{Poisson case with known background}
\label{sec:rstar_simple}

For the known-background counting experiment, the auxiliary statistic for a
tested value of $s$ is~\cite{DavisonEtAl2006}

\begin{equation}
  \label{eq:q_simple_gen}
  u(s) = \sqrt{n}\,\ln\!\frac{n}{s+b} \;.
\end{equation}

\noindent This determines the higher-order-corrected
discovery statistic $q_0^\ast$ for the observed count $n$. Since the
known-background counting experiment involves a discrete Poisson
variable, the continuity correction introduced in
Sec.~\ref{sec:continuity} is applied when evaluating $q_0^\ast$.

To summarise the ingredients for calculating the observed and median
significances with both the Barndorff-Nielsen $r^{\ast}$ and continuity
corrections, we first write the observed significance as
$Z = \sqrt{q_0^{\ast}}$, where $q_0^{\ast}$ is defined in terms of
$r^{\ast}(0)$ by Eq.~(\ref{eq:q0star}). The statistic $r^{\ast}(0)$ is found
from Eq.~(\ref{eq:rstar}) with $s=0$,

\begin{equation}
\label{eq:rstar0fromrandu}
    r^{\ast}(0) = r(0) + \frac{1}{r(0)} \ln \left| \frac{u(0)}{r(0)} \right| \;,
\end{equation}

\noindent where 

\begin{equation}
    r(0) = \operatorname{sign}(n-b)
    \sqrt{2\left[ n \ln \frac{n}{b} + b - n \right]} \;,
\end{equation}

\noindent and from Eq.~(\ref{eq:q_simple_gen}), 

\begin{equation}
    u(0) = \sqrt{n} \ln \frac{n}{b} \;.
\end{equation}

\noindent For $n=0$, taking the appropriate limits 
for products with logarithms corresponds to 
$n\ln n \to 0$ and  $\sqrt{n}\ln n \to 0$. 
In this case one has $u(s)=0$ and the logarithmic $r^\ast$ 
adjustment is undefined.  For $n=0$ we therefore adopt $r^\ast(s)=r(s)$ 
so that the statistic remains well defined numerically.

Finally, we apply the continuity correction by evaluating the statistic at \(n-\tfrac{1}{2}\). For the median significance, \(n\) is replaced by the Asimov value \(s+b\), so that the continuity-corrected statistic is evaluated at \(s+b-\tfrac{1}{2}\). 

\begin{figure}[b]
  \centering
  \includegraphics[width=\columnwidth]{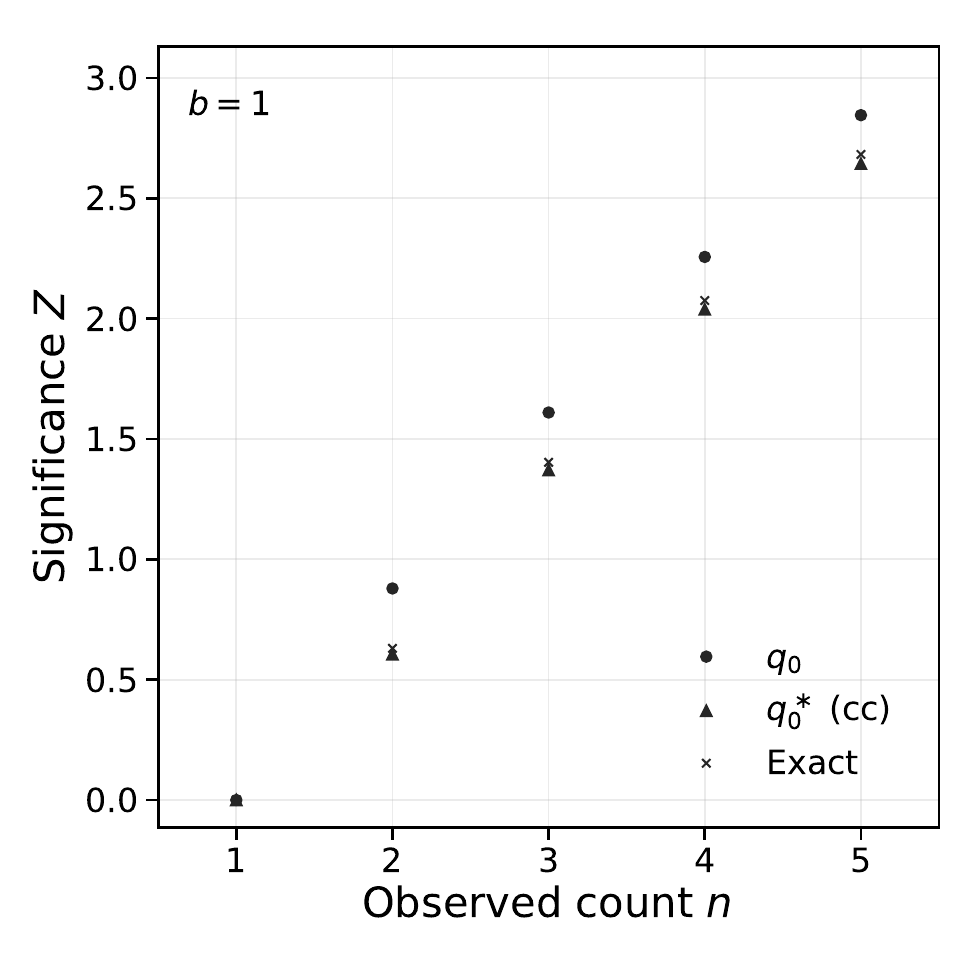}
  \caption{Observed discovery significance $Z = \Phi^{-1}(1-p)$ versus
    observed count $n$ for the known-background counting experiment with
    $b = 1$, testing $s = 0$. The exact Poisson-tail significance is
    compared with the significances obtained from the first-order statistic
    $q_0$ and the higher-order statistic $q_0^\ast$ with continuity correction.}
  \label{fig:rstar_pval_simple}
\end{figure}

Figure~\ref{fig:rstar_pval_simple} compares, for $b = 1$, the observed
discovery significance $Z$ as a function of the observed count $n$
computed in three ways: from the first-order statistic $q_0$, from the
higher-order statistic $q_0^\ast$ with the continuity correction applied
(labelled ``cc'' in the legend), and from the exact Poisson tail.

As seen in Fig.~\ref{fig:rstar_pval_simple}, for $b = 1$ the 
first-order approximation tends to overestimate the
exact significance, whereas the continuity-corrected higher-order
approximation tracks the exact values closely. If the continuity 
correction is not applied to $q_0^\ast$, its agreement with
the exact significance is found in fact to deteriorate. This 
occurs precisely because 
$q_0^\ast$ accurately approximates the wrong quantity: the 
mid $p$-value rather than
the inclusive Poisson tail relevant for discovery. By contrast, 
applying the continuity correction only to $q_0$ does not produce
an overall improvement; the intrinsic error of $q_0$ exceeds the 
discreteness effect, and the result is an underestimate of 
approximately the same magnitude as the original overestimate.

\subsection{Poisson case with uncertain background}
\label{sec:rstar_onoff}

For the uncertain-background model defined in Eq.~(\ref{eq:onoffL}) and
using the profiled background value $\hat{\hat b}(0)$ from Eq.~(\ref{eq:bhathat0}),
the signed likelihood-ratio root for a test of $s=0$ is 

\begin{equation}
\begin{split}
r(0) ={}&
\operatorname{sign}(\hat{s})
\sqrt{-2\ln\frac{L(0,\hat{\hat{b}}(0))}
                 {L(\hat{s},\hat{b})}}
\\
={}&
\operatorname{sign}(n-\hat{b})
\Biggl\{-2\Biggl(
n\ln\left[\frac{n+m}{(1+\tau)n}\right]
\\
&\qquad{}
+m\ln\left[
\frac{\tau(n+m)}{(1+\tau)m}
\right]
\Biggr)\Biggr\}^{1/2}\;.
\end{split}
\end{equation}

\noindent The auxiliary statistic $u(s)$ for a general test of $s$ is

\begin{equation}
\label{eq:q_onoff_gen}
\begin{split}
u(s) ={}&
\sqrt{nm}
\left[
\frac{n}{(s+\hat{\hat b}(s))^2}
+\frac{m}{\hat{\hat b}(s)^2}
\right]^{-1/2}
\\
&\times
\Biggl[
\frac{1}{\hat{\hat b}(s)}
\ln\frac{n}{s+\hat{\hat b}(s)}
\\
&\qquad{}
-\frac{1}{s+\hat{\hat b}(s)}
\ln\frac{m}{\tau\hat{\hat b}(s)}
\Biggr]\;,
\end{split}
\end{equation}
%
%
\noindent which for $s=0$ reduces to
\begin{equation}
  \label{eq:q_onoff}
  u(0) = \sqrt{\frac{nm}{n+m}}\,\ln\frac{n\tau}{m} \;.
\end{equation}

\noindent When either $n=0$ or $m=0$, the terms in $r(0)$ are interpreted
through their continuous limits, while the auxiliary statistic is assigned its
limiting value $u(0)=0$, since $\sqrt{x}\ln x\to0$. At these sample-space
boundaries, the logarithmic $r^\ast$ adjustment is undefined, and we again
adopt $r^\ast(0)=r(0)$.

Following the same steps as in the known-background case, 
the $r^{\ast}$ correction is found by using the $r(0)$
and $u(0)$ given above in Eq.~(\ref{eq:rstar0fromrandu}),
$q_0^{\ast}$ again is obtained from Eq.~(\ref{eq:q0star}), and 
the significance is $Z = \sqrt{q_0^{\ast}}$.  

As discussed in Sec.~\ref{sec:continuity}, no continuity correction is
applied to the two-dimensional data pair $(n,m)$. To find the median
discovery significance, we evaluate the expressions, as before, using the
Asimov data set $n \rightarrow s+b$ and $m \rightarrow \tau b$. To express
the result in terms of $s$, $b$ and $\sigma_b$, we eliminate $\tau$ using
$\tau = b/\sigma_b^2$ (see Eq.~(\ref{eq:sigma2b})), where $\sigma_b$ is
the standard deviation of the estimated background.

\begin{figure}[t]
  \centering
  \includegraphics[width=\columnwidth]{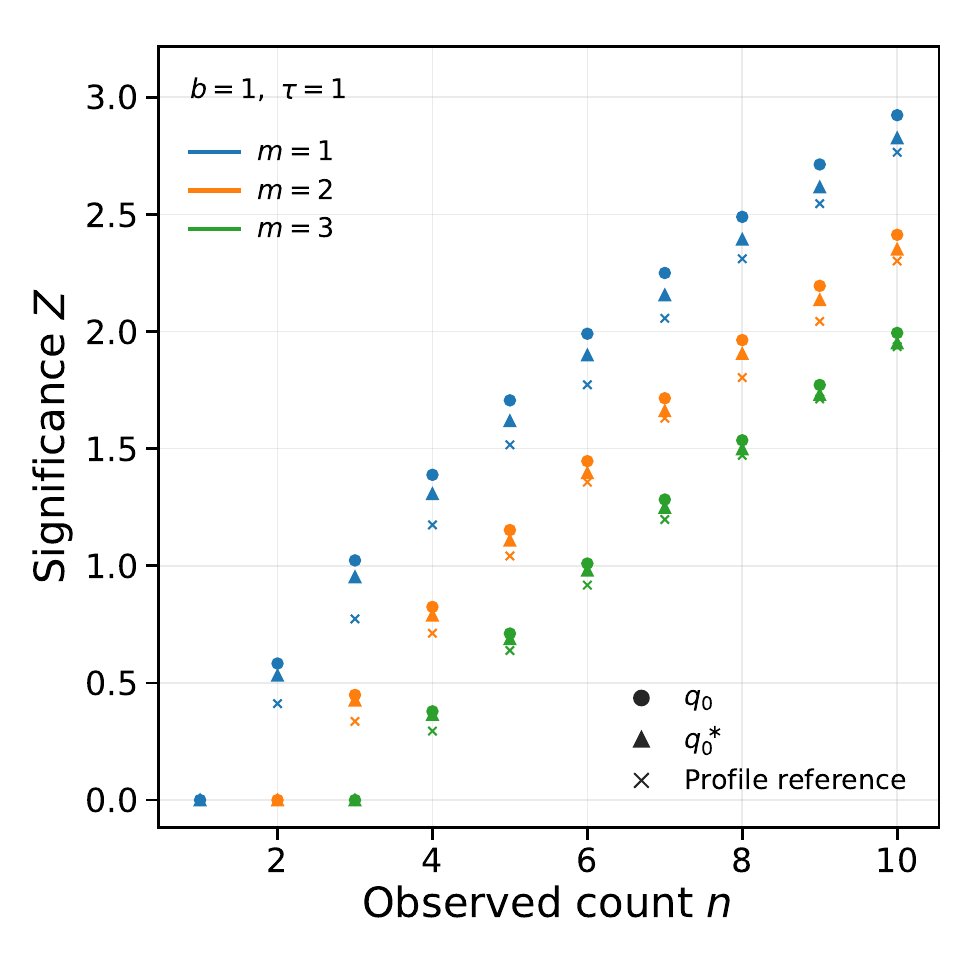}
  \caption{Observed discovery significance $Z$ versus primary count
    $n$ for the uncertain-background model with $b=1$ and $\tau=1$, testing
    $s=0$. Three values of the control count $m$ are shown. The profile reference
    significance is compared with the significances obtained from $q_0$ and
    $q_0^\ast$.}
  \label{fig:rstar_pval_onoff}
\end{figure}

Figure~\ref{fig:rstar_pval_onoff} compares the observed significance
computed using the corrected discovery test statistic $q_0^\ast$ with the
first-order result based on $q_0$ and with the reference significance obtained
using the profile-likelihood construction discussed in Sec.~\ref{sec:bunknown},
for $b=1$ and $\tau=1$. Three representative values of the control count
$m$ near its expected value, $E[m] = \tau b$, are shown. In the discovery
tail, the significance obtained from $q_0^\ast$ provides a better approximation
to the MC reference significance than that obtained from $q_0$. The improvement
is largest when both expected counts, $E[n]=s+b$
and $E[m]=\tau b$, are small. For larger effective yields, the two
approximations converge to the MC reference result.

\subsection{Higher-order corrections to the median 
discovery significance}
\label{sec:rstar_asimov}

Evaluating $q_0^\ast$ with the Asimov data set involves two independent
levels of approximation.  The first is the Asimov approximation itself: 
the median discovery significance is estimated by evaluating $Z$ at 
the expected data $n = s + b$ (and $m = \tau b$ for the
uncertain-background problem) in place of computing the true median of the
test-statistic distribution. The second is the choice of test-statistic
approximation at that fixed data set: the first-order statistic $q_0$ or the
corrected statistic $q_0^\ast$. The $q_0^\ast$ correction addresses only the second
level, providing a more accurate significance for the specific Asimov
data point.

\begin{figure}[t]
  \centering
  \includegraphics[width=\columnwidth]{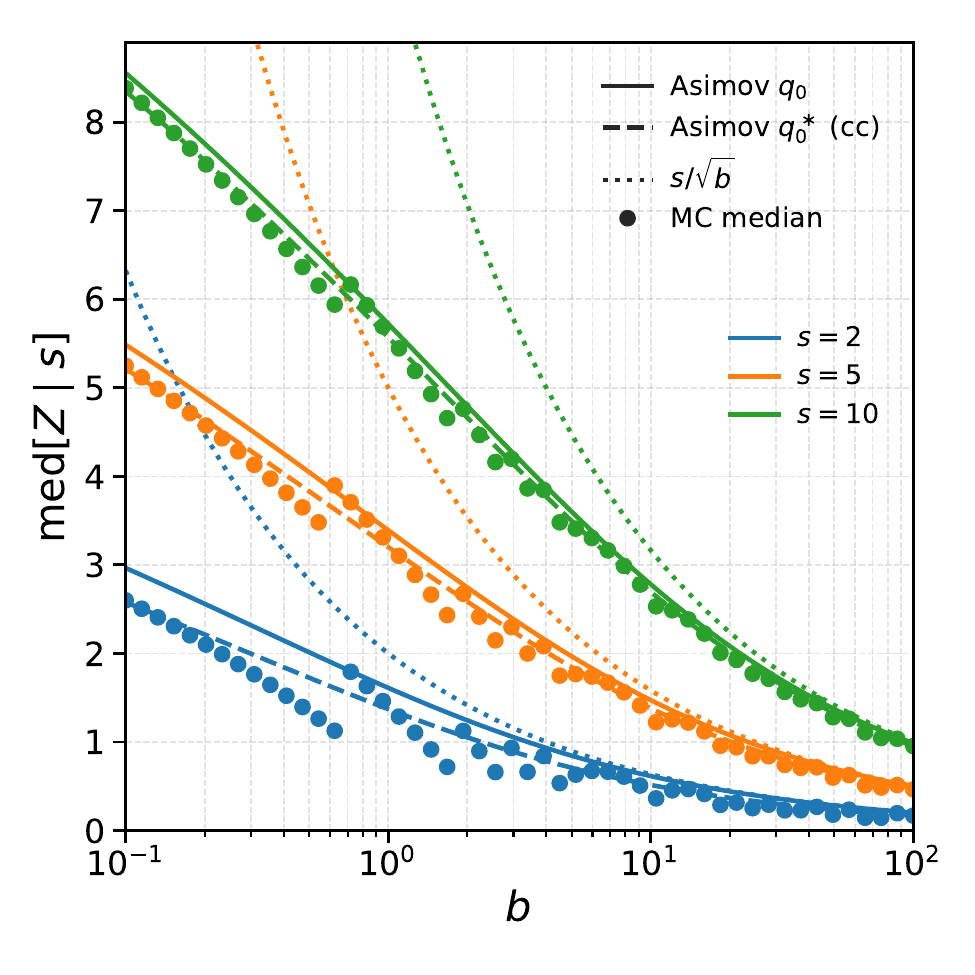}
  \caption{Median discovery significance $\mbox{med}[Z | s]$ as a function
    of the expected background $b$ for the known-background counting experiment
    with $s=2,5$ and $10$. The curves show the Asimov significance
    $Z_{\rm A}$ obtained from $q_0$, the limiting expression $s/\sqrt{b}$,
    and the MC estimate of $\mbox{med}[Z | s]$ (same as in
    Fig.~\ref{fig:medsig}). The plot also includes the $Z_{\rm A}$ prediction
    based on $q_0^{\ast}(cc)$ with higher-order and continuity
    corrections (the latter is applied to the continuous Asimov value $n=s+b$).}
  \label{fig:rstar_medsig_simple}
\end{figure}

Figure~\ref{fig:rstar_medsig_simple} shows the results for the known-background
counting experiment with $s=2,5$ and $10$. For $q_0^\ast$, the continuity correction described in
Sec.~\ref{sec:continuity} is applied to the continuous Asimov value, $n\rightarrow n-\tfrac12$.
The Asimov significance $Z_{\rm A}$ obtained from $q_0^\ast$ agrees better
with the MC estimate of $\mbox{med}[Z | s]$ than that obtained from $q_0$
across the full range of $b$, reflecting the improved significance
approximation at the Asimov point. The plot shows how combining the higher-order approximation to the
test-statistic distribution with the continuity correction allows the
$Z_{\rm A}$ prediction based on $q_0^\ast$ to interpolate the MC values near
the centres of their discrete steps.

\begin{figure*}[p]
  \centering
  \includegraphics[width=0.84\textwidth]{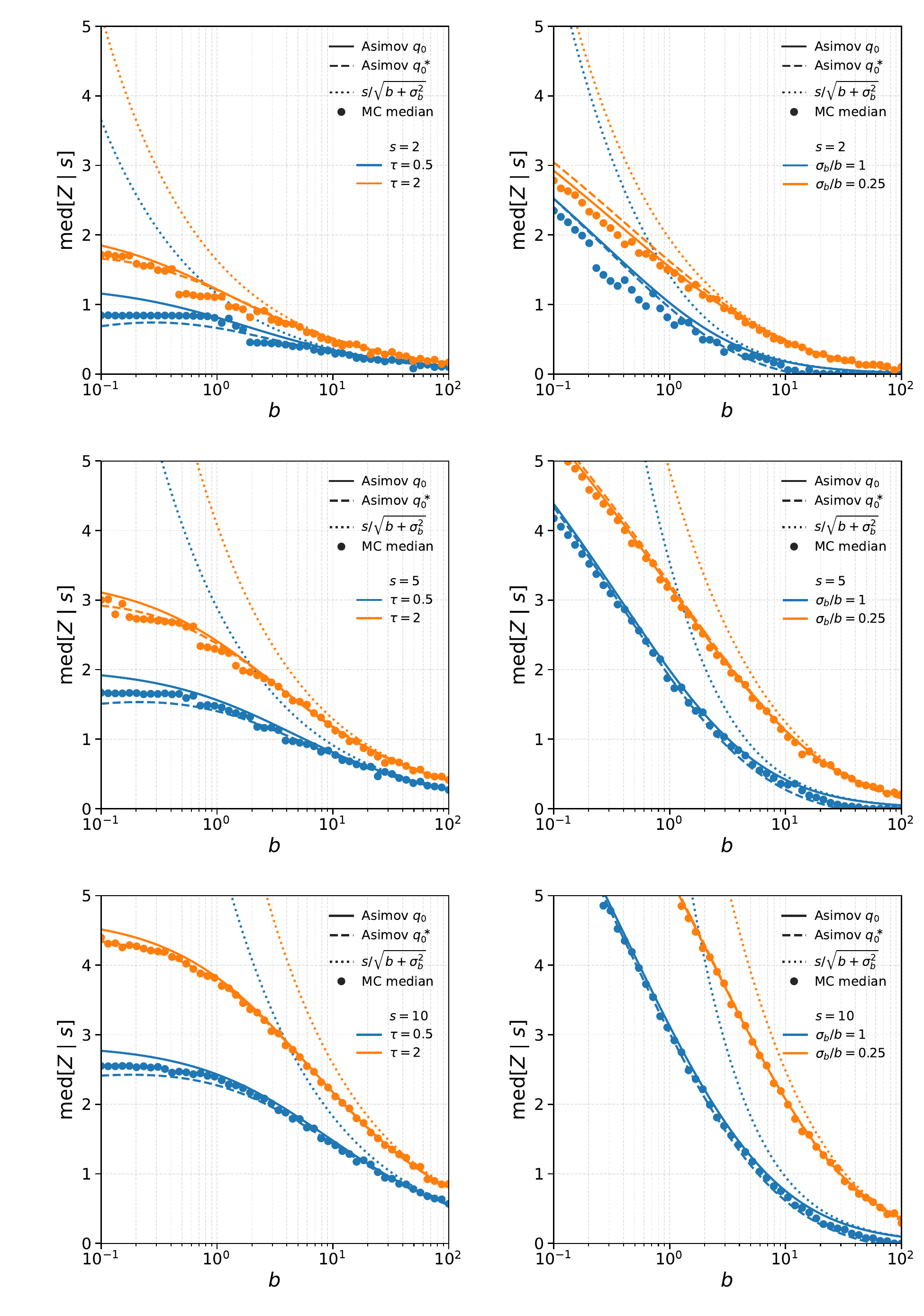}
  \caption{Median discovery significance $\mbox{med}[Z | s]$ versus $b$ for
    the uncertain-background model with $s=2$ (top), $s=5$ (middle) and $s=10$
    (bottom).  The left column fixes $\tau=0.5$ or $2$, and the right
    column fixes $\sigma_b/b=0.25$ or $1$. The plot shows the Asimov
    significance $Z_{\rm A}$ obtained from $q_0$, the limiting formula
    $s/\sqrt{b+\sigma_b^2}$, and the MC estimate of $\mbox{med}[Z | s]$
    (same as in Fig.~\ref{fig:medsigsys}). In addition, the curves include
    the $Z_{\rm A}$ prediction based on $q_0^{\ast}$, which improves the
    agreement with the MC estimate for lower event yields. MC points that
    reach the $5\sigma$ toy-resolution ceiling are omitted.}
  \label{fig:rstar_medsig_onoff}
\end{figure*}

For the uncertain-background problem, Fig.~\ref{fig:rstar_medsig_onoff} 
shows the median discovery significance in the same manner as in
Fig.~\ref{fig:medsigsys}:  the left-hand panels correspond to fixed $\tau$
and the right-hand ones to fixed $\sigma_b/b$; the top, middle and
bottom rows use $s=2, 5$ and $10$, respectively. 

For most parameter values shown in the plots, the higher-order correction 
to $\operatorname{med}[Z\mid s]$ is seen to provide additional improvement 
in the agreement with the MC estimates, especially at low $b$ for fixed $\tau$ (left-hand plots) and at high $b$ for fixed $\sigma_b/b$ (right-hand plots). An interesting exception occurs 
at small $b$ and small $\sigma_b/b$.  In this limit, 
$E[m] = 1/(\sigma_b/b)^2$ can
be large compared with the expected primary count, $E[n]=s+b$, so the
control measurement effectively fixes the background; this regime is 
illustrated by the $s=2$ examples in Fig.~\ref{fig:rstar_medsig_onoff}. 
The problem therefore approaches the known-background case, with $u(0)$
in Eq.~(\ref{eq:q_onoff}) reducing to Eq.~(\ref{eq:q_simple_gen}). The
continuity correction is correspondingly most relevant in this limit, 
where the discreteness of $n$ is no longer smeared by the auxiliary 
measurement.  Since we cannot, however, apply a continuity correction
with two discrete measurements $n$ and $m$, $q_0^\ast$ 
loses some accuracy as this limit is approached, as seen
in Fig.~\ref{fig:rstar_medsig_onoff} for $s=2$ and $\sigma_b/b = 0.25$
at low $b$.

\section{Conclusions}
\label{sec:conclusions}

A compact expression for the expected discovery significance is given
by Eq.~(\ref{eq:onoffmedZ2}), which applies to a counting experiment in
which the expected background is constrained by a Poisson control
measurement.  This agrees with the naive formula 
Eq.~(\ref{eq:soverrootbsys}) in the limit of $b \gg s$
and small background uncertainty.
The numerical studies shown here indicate, however, that for small
values of background rate and uncertainty, this limiting formula can
overestimate the expected discovery significance, whereas the
full formula remains in very good agreement with exact Monte Carlo
results.

The same comparisons also show where higher-order corrections become
important. The accuracy of both the observed and the median
discovery significance can be further improved by the higher-order
Barndorff-Nielsen $r^\ast$ correction, here specialised to the discovery
test through the corrected statistic $q_0^\ast$ of Eq.~(\ref{eq:q0star}).
For the known-background case, using $q_0^\ast$  together
with a continuity correction reduces the discrepancy between the
asymptotic and exact observed significances to a very small level, even
for expected counts as low as one. This also shows that, in discrete 
problems,  the accuracy of $q_0^\ast$
can depend not only on the higher-order correction itself, but also on
matching the continuous approximation to the appropriate definition of
the discrete tail probability. For the
uncertain-background case, $q_0^\ast$ still provides a clear improvement
over the first-order approximation when both the expected on- and
off-region counts, $E[n]=s+b$ and $E[m]=\tau b$, are small. However, 
the absence of a canonical continuity correction when more
than one discrete count is involved can limit the achievable accuracy
in the lowest-count regions, and should be kept in mind when applying
$q_0^\ast$ to discrete models.

\backmatter

\section*{Acknowledgments}  The authors thank 
colleagues in the ATLAS Collaboration including Kyle Cranmer, Eilam Gross, and Ofer Vitells for valuable input relating to this work.  
We are grateful to the UK Science and Technology Facilities Council for 
its support.

\section*{Declarations}

\bmhead{Data availability}

No experimental data were analysed for this study. The numerical
results and figure data can be reproduced using the code and configurations
provided in the repository cited in the Code availability statement.

\bmhead{Code availability}

The code and configurations used to generate the numerical results and
figures are openly available in the Count Significance repository,
release~\texttt{v0.2.0}, at
\url{https://github.com/EnzoCanonero/count-significance/releases/tag/v0.2.0}.
The corresponding Python package is available from PyPI as
\texttt{count-significance} version~\texttt{0.2.0} at
\url{https://pypi.org/project/count-significance/0.2.0/}.

\bibliography{references}

\end{document}